\documentclass[prl,aps,floatfix,superscriptaddress,showpacs,showkeys,amsmath,amssymb,twocolumn,]{revtex4-1}
\usepackage{amsthm}
\usepackage{amsfonts}
\usepackage{amstext}
\usepackage{enumitem}
\usepackage{graphicx}
\usepackage{tabularx}
\usepackage{float}
\usepackage{multirow}
\usepackage{color}
\usepackage{lipsum}
\usepackage[dvipsnames]{xcolor}
\usepackage{tkz-kiviat,numprint} 
\usetikzlibrary{arrows}
\usetikzlibrary{plotmarks}
\usetikzlibrary{shapes}
\usepackage{afterpage}
\usepackage{xspace}
\usepackage{rotating}
\usepackage{upgreek}
\usepackage{morefloats}
\usepackage{url}
\usepackage[caption=false]{subfig}
\usepackage{color}
\usepackage{mje}

\def\Bid{{\mathchoice {\rm {1\mskip-4.5mu l}} {\rm
{1\mskip-4.5mu l}} {\rm {1\mskip-3.8mu l}} {\rm {1\mskip-4.3mu l}}}}

\newcommand{\Param}{\ensuremath{\boldsymbol{\Omega}}\xspace}
\newcommand{\Func}[1]{\ensuremath{\mathrm{W}_{#1}\left(\Param\right)}\xspace}
\newcommand{\Parity}{\ensuremath{\hat \Pi}\xspace}
\newcommand{\Dim}{\ensuremath{\mathfrak{D}}\xspace}

\newcommand{\OperBare}{\ensuremath{\hat{\Delta}}\xspace}
\newcommand{\Oper}{\ensuremath{\OperBare\left(\Param\right)}\xspace}

\newcommand{\DO}{{\hat{\rho}}}

\newcommand{\QSERG}{Quantum Systems Engineering Research Group, Department of Physics, Loughborough University, Leicestershire LE11 3TU, United Kingdom}

\begin{document}


\title{Wigner Functions for Arbitrary Quantum Systems}
\author{Todd Tilma}\email{tilma.t.aa@m.titech.ac.jp}
\affiliation{Tokyo Institute of Technology, 2-12-1 Ookayama, Meguro-ku, Tokyo 152-8550, Japan}
\author{Mark J. Everitt}\email{m.j.everitt@physics.org}
\affiliation{\QSERG}
\author{John H. Samson}
\affiliation{\QSERG}
\author{William J. Munro}
\affiliation{NTT Basic Research Labs, NTT Corporation, 3-1 Morinosato-Wakamiya, Atsugi, Kanagawa 243-0198, Japan}
\affiliation{National Institute of Informatics, 2-1-2 Hitotsubashi, Chiyoda-ku, Tokyo 101-8430, Japan}
\author{Kae Nemoto}
\affiliation{National Institute of Informatics, 2-1-2 Hitotsubashi, Chiyoda-ku, Tokyo 101-8430, Japan}
\date{\today}

\begin{abstract}
The possibility of constructing a complete, continuous Wigner function for any quantum system has been a subject of investigation for over 50 years. 
A key system that has served to illustrate the difficulties of this problem has been an ensemble of spins. 
Here we present a general and consistent framework for constructing Wigner functions exploiting the underlying symmetries in the physical system at hand. 
The Wigner function can be used to fully describe any quantum system of arbitrary dimension or ensemble size.
\end{abstract}

\maketitle


Out of all available choices, one can argue that the Wigner function~\cite{Wigner1932} presents the most natural phase-space representation of quantum mechanics~\cite{Wigner1984}.
The main advantage is that it simultaneously retains the intuitiveness with respect to classical phase-space while rendering clearly, important quantum information concepts - leading to the now iconic Wigner function for macroscopically distinct superposition of states (Schr\"odinger cat states)~\cite{Deleglise:2008gt}.  
In this regard the Wigner function possesses a unique advantage over other representations (such as the $P$~\cite{PhysRevLett.10.277,PhysRev.131.2766} and $Q$~\cite{Husimi1940,0305-4470-34-10-309} functions). 
Even though all these are quasiprobability distribution functions, the Wigner function's marginals are easily linked to amplitudes of a given representation, and its equations of motion are closely and intuitively relatable to the classical ones for the same system~\cite{Wigner1984}.
These properties are further augmented by a transparent connection to the quantum-classical transition where solutions to the classical Liouville equation can be recovered as the action becomes large with respect to a Planck cell~\cite{PhysRevLett.80.4361}. 
Indeed, it is possible to reformulate much of quantum mechanics in pahse space~\cite{Curtright:2012ka}.

Despite the merits of the Wigner function representation, and its successful application in quantum optics~\cite{Scully1997,BillMunro1}, it has not been more widely applied to other systems as finding a consistent approach to generating Wigner functions for arbitrary, finite dimensional systems has proved challenging.
For example, Wigner functions for finite-dimensional systems have been developed~\cite{Wootters19871,PhysRevA.53.2998,VOURDAS1997367,PhysRevA.65.062309,PhysRevA.70.062101}, but their definition is restricted to a subset of discrete state-spaces.
Furthermore, only gradual progress has been made in the development of continuous state-space Wigner functions representing finite dimensional systems~\cite{Agarwal1981,Shibata1976,PhysRevA.59.971,Braunstein-1004.5425,Klimov-1008.2920,Wolf6247,Luis-052112,Luis-495302,Klimov055303,TilmaKae1,PhysRevA.86.062117}. 
These approaches also come with their own set of restrictions: the representation space is restricted to the symmetric subspace where the Bloch sphere can be constructed, or the representation space is expanded to support the entire Hilbert space at the cost of distorting the properties of the state or states being represented.  
It is clear therefore, that the most appropriate Wigner function for an arbitrary quantum system should be one that is a complete representation, which preserves the quantum properties of the system in an intuitive way, yet is consistent and comparable with continuous variable cases from quantum optics.  

In this Letter, based on the original Wigner function for continuous variable systems, we propose an alternative method for computing Wigner functions that addresses all these issues and thus provides a pathway to the formulation of intuitively analogous, easy to calculate, complete Wigner functions for arbitrary quantum systems.
As proof of principle, we present examples of Wigner functions that are currently of importance in both quantum information and atomic/molecular/optical physics.

The standard form of the Wigner function describing how to transform a Hilbert space operator $\hat{\rho}$ to a classical phase-space function $\mathrm{W}_{\hat{\rho}}(\mathbf{q},\mathbf{p})$~\cite{Weyl1927,Moyal1949,Ozizmir1967,Leaf1968-1,Leaf1968-2}, is
\bel{Weyl-Wigner-ND1}
\mathrm{W}_{\hat{\rho}}(\mathbf{q},\mathbf{p}) = \biggr(\frac{1}{ 2 \pi \hbar}\biggl)^{n} \int_{-\boldsymbol{\infty}}^{+\boldsymbol{\infty}} \ud \mathbf{z} \, \bra{\mathbf{q}-\frac{\mathbf{z}}{2}} \hat{\rho} \ket{\mathbf{q}+\frac{\mathbf{z}}{2}} \, e^{i \mathbf{p} \cdot \mathbf{z}/\hbar}, 
\ee 
where $\mathbf{q}=[q_1,q_2,\ldots,q_n]$ and $\mathbf{p}=[p_1,p_2,\ldots,p_n]$ are $n$-dimensional vectors representing the classical phase-space position and momentum values, $\mathbf{z}=[z_1,z_2,\ldots,z_n]$, $\hbar$ is Planck's constant, and with normalization
\bel{Weyl-Wigner-ND2}
\int_{-\boldsymbol{\infty}}^{+\boldsymbol{\infty}}\ud \mathbf{q} \int_{-\boldsymbol{\infty}}^{+\boldsymbol{\infty}}\ud \mathbf{p} \, \mathrm{W}_{\hat{\rho}}(\mathbf{q},\mathbf{p})  =  \Trace{\hat{\rho}} = 1.
\ee
It is well known that this can be also written in terms of the displacement ($\OpD$) and parity ($\Parity$) operators according to:
\bel{WignerStart}
\mathrm{W}_{\hat{\rho}}(\Param)=\biggr(\frac{1}{ \pi \hbar}\biggl)^{n} \Trace{ \hat{\rho} \, \OpD(\Param) \Parity \OpD^\dag(\Param) }
\ee
where $\Param$ is any full parametrization of the phase space such that $\OpD$ and $\Parity$ are defined in terms of coherent states $\OpD(\Param) \ket{0} = \ket{\Param}$ and $\Parity \ket{\Param} = \ket{-\Param}$~\cite{PhysRevA.50.4488,PhysRevA.48.2479}.
In this situation, the displacement operator $\OpD$ is often parametrized in terms of position and momentum coordinates or eigenvalues of the annihilation operators.
The question then is, especially for composite quantum systems, \emph{can this displaced parity operator approach be generalized to other, especially spin, systems?}
In other words, we want an equation of the form of~\Eq{WignerStart} but for finite-dimensional, continuous variable, composite quantum systems.

We will follow the approach of Brif and Mann~\cite{PhysRevA.59.971} by considering a distribution $W_{\hat{\rho}}(\Omega)$ over a phase space defined by the parameters $\Omega$ to be a Wigner function of a Hilbert space operator $\hat{\rho}$ if there exists a kernel $\hat{\Delta}(\Omega)$ that generates $W_{\hat{\rho}}(\Omega)$ according to the generalized Weyl rule $W_{\hat{\rho}}(\Omega)=\Trace{\hat{\rho}\hat{\Delta}(\Omega)}$ and which also satisfies the following restricted version of the Stratonovich-Weyl correspondence:  
\begin{enumerate}[label=\sffamily \footnotesize \upshape S-W.\arabic*]
\item\label{D1} The mappings $\Func{\DO}=\Trace{\DO \, \Oper}$ and $\DO = \int_{\Param} \Func{\DO} \Oper \ud \Param$ exist and are informationally complete. Simply put, we can fully reconstruct $\DO$ from $\Func{\DO}$ and vice versa~\footnote{For the inverse condition, an intermediate linear transform may be necessary.}.
\item\label{D2} $\Func{\DO}$ is always real valued which means that $\Oper$ must be Hermitian. 
\item\label{D3}  $\Func{\DO}$ is standardized so that the definite integral over all space $\int_{\Param} \Func{\DO} \ud \Param = \Tr{\DO}$  exists and $\int_{\Param} \Oper \ud \Param =\Bid$.
\item\label{D4} Unique to Wigner functions, $\Func{\DO}$ is self-conjugate; the definite integral $\int_{\Param} \Func{\DO'}\Func{\DO''} \ud \Param= \Trace{\DO' \DO''} $ exists. 
This is a restriction of the usual Stratonovich-Weyl correspondence. 
\item\label{D5} Covariance:
Mathematically, any Wigner function generated by ``rotated'' operators $\hat{\Delta}(\Param^{\prime})$ (by some unitary transformation $\OpU$) must be equivalent to rotated Wigner functions generated from the original operator ($\hat{\Delta}(\Param^{\prime}) \equiv \OpU \Oper \OpU^{\dagger}$) - \textit{i.\ e.\ }if $\DO$ is invariant under global unitary operations then so is $\Func{\DO}$.
\end{enumerate}
We note that the kernel operator $\Oper$ and the set of coordinates $\Param$ are not unique under the conditions for a phase-space function to be a Wigner function.

For continuous systems \Eq{WignerStart} shows the kernel operator $\Oper$ to be proportional to $\OpD(\Param) \Parity \OpD^\dag(\Param)$ with the parameters $\Param=\{\mathbf{q},\mathbf{p}\}$.  
For other systems, it is essential for the kernel operator (and the set of coordinates) to be chosen in order to reflect the symmetries of the physical system of interest. 
As an example, we start with Definiton~\ref{D1} and attempt to recreate an analogous equation to~\Eq{WignerStart} for a single, two-level, quantum system.
In this case, $\Parity$ has analogous properties to $\hat{\sigma}_z$: acting as a $\pi$-rotation on a two-level quantum system about the $z$-axis of the Bloch sphere in the Pauli representation.
Similarly, the $\SU{2}$ rotation operator, $\OpU_2^{[2]}(\theta,\phi,\Phi)=e^{\ui \hat{\sigma}_z \phi}e^{\ui \hat{\sigma}_y \theta}e^{\ui \hat{\sigma}_z \Phi}$, is analogous to the displacement operator $\OpD$ in that $\OpU_2^{[2]}(\theta,\phi,\Phi)$ ``displaces'' a two-level quantum state along the surface of the Bloch sphere.
Where necessary, we use bracketed superscripts $[D]$ to represent the $D \times D$ matrix size of the operator, and numerical subscripts $\Dim$ to denote the operator's Special Unitary (SU) group structure. 

In order to obtain a Wigner function from the above, we are motivated to take the rotated $\hat{\sigma}_z$ operator as the displaced parity operator for the two-level system and impose the self-conjugate Stratonovich-Weyl correspondence~\cite{PhysRevA.50.4488}.  
This argument leads to the following expression~\cite{TilmaKae1,Problem-Cite}
\bel{eq:WF1s}
\OperBare^{[2]}(\theta,\phi)=\Half \left[\Id^{[2]} -\sqrt{3}\left(\OpU_2^{[2]} \hat{\sigma}_z(\OpU_2^{[2]})^\dag\right) \right]
\ee
where the Euler angles $(\theta,\phi)$ parametrizing the representation space are set by the parametrization of the rotation operator $\OpU_2^{[2]}(\theta,\phi,\Phi)$.  
Using the invariance of the $2 \times 2$ identity $\Id^{[2]}$ under $\OpU_2^{[2]}$ we have
\bel{TheSpinParity}
\hat \Pi^{[2]}  = \Id^{[2]} -\sqrt{3}\, \hat{\sigma}_z 
\ee
such that
\bel{eq:WF1sF}
\OperBare^{[2]}(\theta,\phi)=  \Half \left[\OpU_2^{[2]} \hat \Pi^{[2]}(\OpU_2^{[2]})^\dag \right].
\ee
It is clear that this operator is Hermitian, and that with the correct $\ud \Param$ (for our discussions, the Haar measure given in~\cite{Tilma2}) satisfies all the requirements of our restricted Stratonovich-Weyl correspondence.
As the spin-parity $\hat{\Pi}$ is an observable and the displacement-rotation $\OpU$ operators are easily realizable quantum operations then, as for optical systems~\cite{PhysRevLett.78.2547,PhysRevA.60.674}, direct reconstruction of our Wigner function should be possible.
For example, it should be possible to set up solid-state-based experiments to directly measure these spin-based Wigner functions.

We can use~\Eq{TheSpinParity} and~\Eq{eq:WF1sF} as a starting point to generalize the construction of the kernel operator $\Oper$.  
To do this, we focus on the symmetries in the physical systems in question. 
We start with a quantum system that is a collection of $k$ distinct states, each being parametrized by a $\SU{n_i}$ spin representation of dimension $d_i$, such that the system size is $D = d_1 \times d_2 \times \cdots \times d_k$ and $\Dim = n_1 \times n_2 \times \cdots \times n_k$.
The full system can then be parametrized by the appropriate $D$-dimensional representation of $\SU{\Dim}$.
From this, the key to formulating an appropriate kernel is clear.
The spin parity operator $\hat \Pi$ needs to address the overall symmetry of the total system, which means it must be an element of the algebra $\mathsf{su}(D)$.
For our work, such an element will be defined using the formalism given in~\Eq{TheSpinParity} by using the last of the generalized Gell-Mann matrices, $\hat{\Lambda}_i$~\cite{Greiner}, which, as $\hat{\sigma}_z$ is $\hat{\Lambda}_{3}$ in $\SU{2}$, is a natural extension of the case considered in~\Eq{eq:WF1s}.

The previous argument leads us to propose that spin Wigner functions can be generated using kernels of the form:
\ba
\label{eq:WFNs}
\OperBare^{[D]}(\Param) & = & \frac{1}{D} \hat{\mathbb{U}}(\Param) \hat \Pi^{[D]}  \hat{\mathbb{U}}^\dag(\Param),\nonumber \\
\hat \Pi^{[D]} & = & \Id^{[D]} -\mathcal N(D) \hat\Lambda_{D^{2}-1}, 
\ea
where the normalization $\mathcal N(D)$ depends on the dimensionality of the Hilbert space and (not denoted here) the choice of $\Param$; $\hat{\mathbb{U}}(\Param) =\bigotimes_{i=1}^k U_{n_i}^{[d_i]}$ and is closed on the parameter space $\Param$ (while we focus on continuous $\Param$ our definition could work in the discrete case too); $\hat{\Lambda}_{D^2-1}$ is a $D \times D$ diagonal matrix wherein the diagonal entries are $\sqrt{\frac{2}{D(D-1)}}$ except for $(\hat{\Lambda}_{D^2-1})_{D, D}=-\sqrt{\frac{2(D-1)}{D}}$~\cite{Greiner}. 
It is clear that the explicit form of $\OperBare^{[D]}(\Param)$ is dependent on the choice of $\hat{\mathbb{U}}$; thus, the question we must address is how to choose such operators so as to satisfy the self-conjugate Stratonovich-Weyl correspondence.

Each choice of $\hat{\mathbb{U}}$, $\hat{\Pi}$, and the parameter space may yield a different Wigner function as long as it satisfies the Stratonovich-Weyl correspondence; hence, a preferred choice of the parameter set should be made to reflect the physical system at hand.
As we focus on spin systems in this Letter, we first consider the standard $\SU{2}$ case and construct the corresponding Wigner function using the above recipe.
A spin-$j$ representation of $\SU{2}$ has been shown to be useful to represent various physical systems such as Bose-Einstein Condensates (BECs)~\cite{Anderson198,PhysRevLett.81.742,0028-0836,1367-2630-8-8-152,nature09887,PhysRevA.55.4318,PhysRevA.90.062132} and spin ensembles in materials~\cite{RevModPhys.76.323}.
Thus, setting $k=1$, $n_1=2$, and $D=d_1=2j+1$ in the definition of $\hat{\mathbb{U}}$ yields the $\SU{2}$ rotations $\OpU_2^{[2j+1]}$.
As such operators can be decomposed with three real parameters $(\phi$, $\theta$, \text{and} $\Phi)$ we have $\OpU_2^{[2j+1]} =e^{\ui \OpJ_3 \phi}e^{\ui \OpJ_2 \theta}e^{\ui \OpJ_3 \Phi}$ where $\OpJ_i$ are the generators of the $[2j+1]$-dimensional representation of $\SU{2}$.
The operators $\Oper$ and $\hat{\Pi}$ are then
\ba
\label{highj}
\OperBare^{[2j+1]}(\theta,\phi) &=& \frac{1}{2j+1}
	\OpU_2^{[2j+1]}
	\Pi^{[2j+1]} 
	(\OpU_2^{[2j+1]})^\dag,
\nonumber \\
\hat \Pi^{[2j+1]}  &=& 
	\Id^{[2j+1]} -
	\mathcal N(2j+1)\hat{\Lambda}_{(2j+1)^2-1}.
\ea
The parameter set, $(\phi, \theta)$, as $\Phi$ makes no contribution, specifies the parameter space for the Wigner function.  
Finally, to obtain an unbiased representation on the parameter space, we take the Haar measure on the parameter space that generates the normalization constant $\mathcal N(2j+1)=\sqrt{{(2j+2)(2j+1)(2j)}/{2}}$. 
\begin{figure}[!t]
    \subfloat[$\ket{\frac{1}{2}}+\ket{-\frac{1}{2}}$]    
        {\includegraphics[width=0.3\linewidth]
        {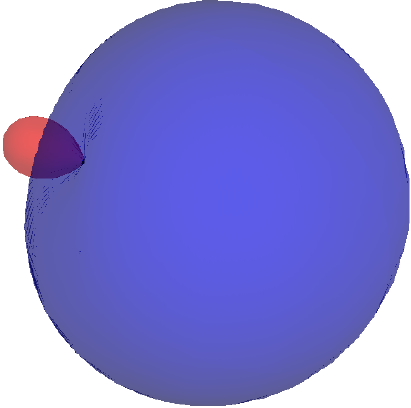}}
        \hfill
    \subfloat[\mbox{$\ket{\frac{3}{2}}+
        \ket{-\frac{3}{2}}$}]
        {\includegraphics[width=0.3\linewidth]
        {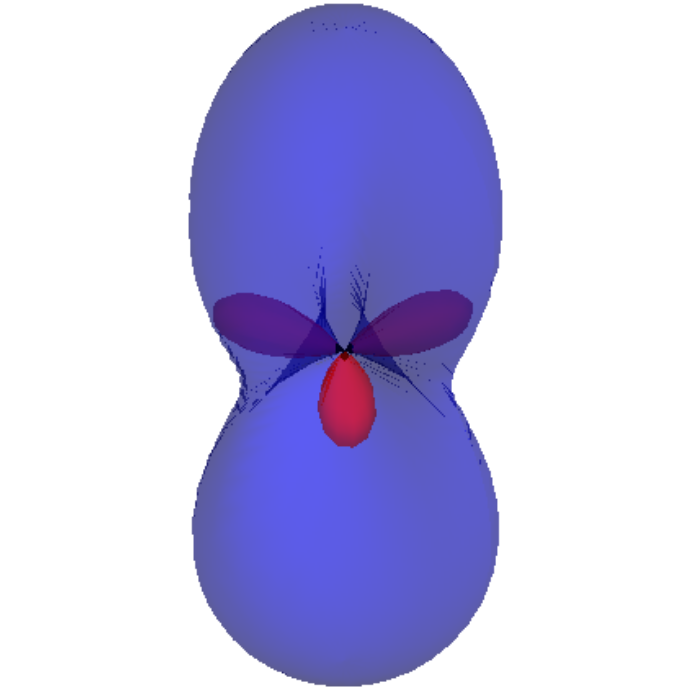}}
        \hfill
    \subfloat[$\ket{\frac{7}{2}}+\ket{-\frac{7}{2}}$]
        {\includegraphics[width=0.3\linewidth]
        {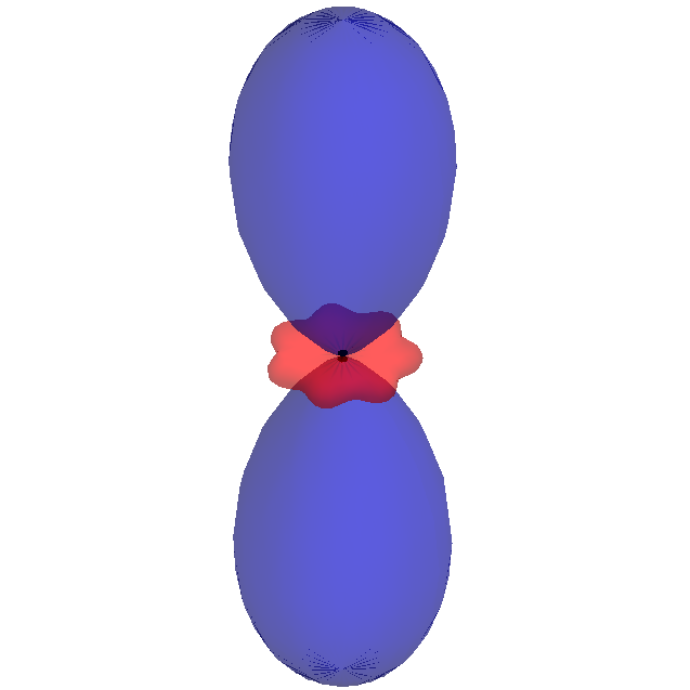}}
    \\
    \subfloat[$\ket{\frac{1}{2}}+\ket{-\frac{1}{2}}$]
        {\includegraphics[width=0.3\linewidth]    
        {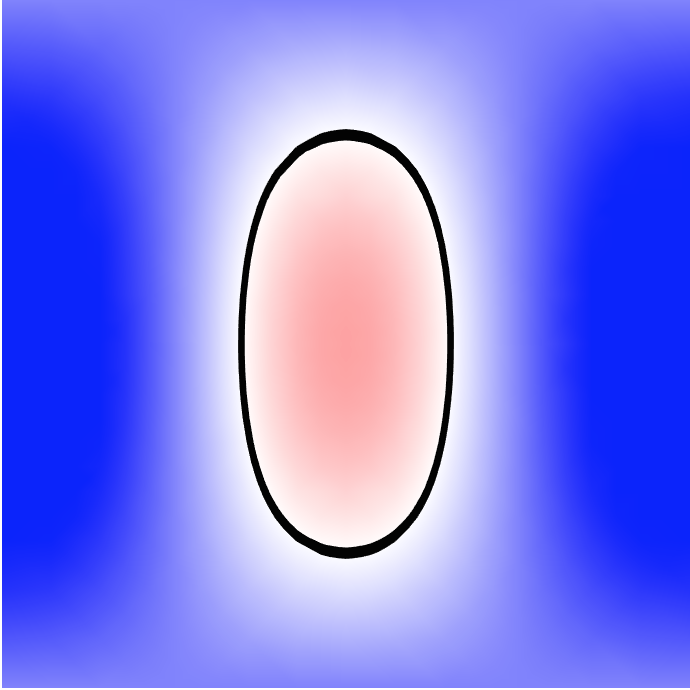}}
    ~
    \subfloat[Bell, \ket{\Phi^+}]    
        {\includegraphics[width=0.3\linewidth]
        {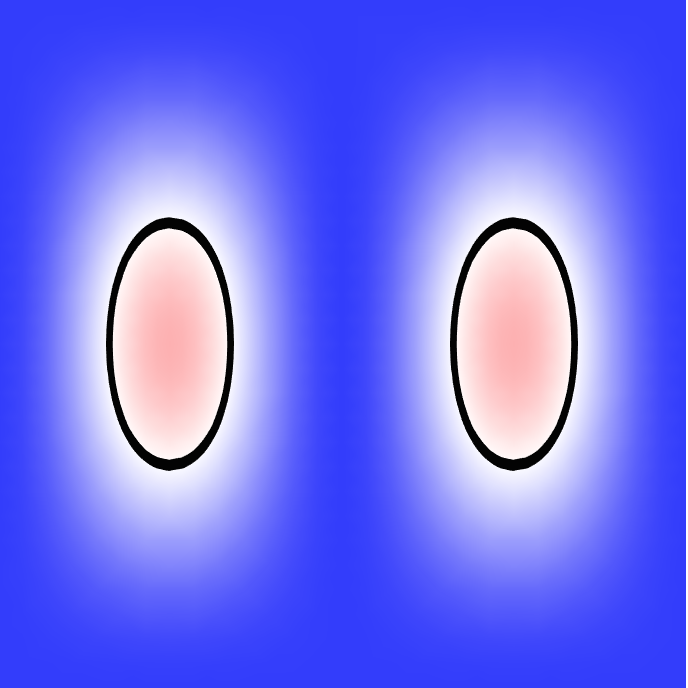}}
    ~
    \subfloat[GHZ state]    
        {\includegraphics[width=0.3\linewidth]
        {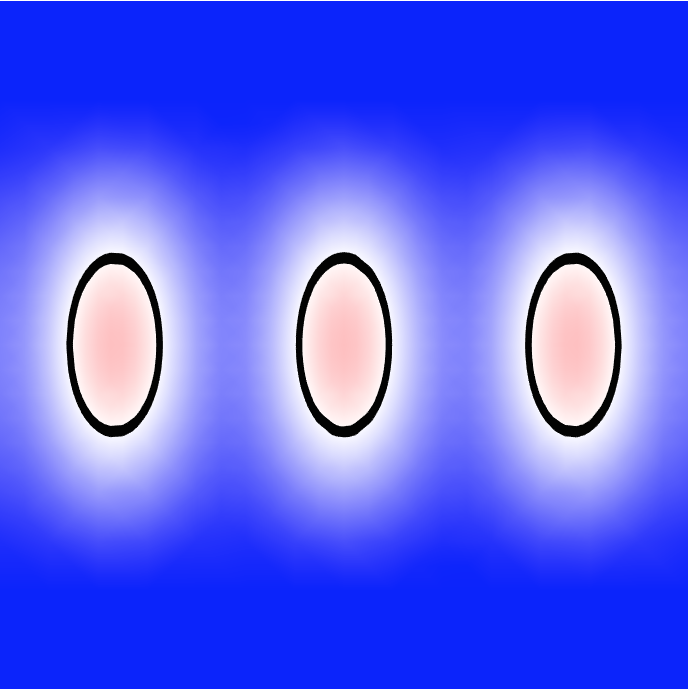}}
\caption{    
    \textbf{(a-c)} Polar plot of the Wigner function using $\OperBare^{[2j+1]}(\theta,\phi)$, 
    as defined in~\Eq{highj} for high spin spaces of 
    dimension  $1/2$ (a),  $3/2$ (b) and $7/2$ (c). Here we 
    have used as examples normalized states of the 
    form $\ket{j, m=j}+\ket{j,m=-j}$ (where we have 
    labeled states in terms of the quantum numbers for 
    $\OpJ^2$ and $\OpJ_z$). Note that there are $2j$ 
    interference terms and that images are not to the 
    same scale. 
    \textbf{(d-f)} Mercator projection of the Wigner function using $\OperBare^{[2^{k}]}(\{\theta_i,\phi_i\})$, as defined in~\Eq{susu}, 
    for a set of $1$ (d), $2$ (e) and $3$ (f) spins/two-level-
    atoms where we have taken the slice $
    \theta_i=\theta$ (as the ordinate from $0$ to $\pi/2$) and $\phi_i=\phi$ 
    (as the abscissa for $0$ to $\pi$).
    For all plots, blue is positive and red negative and black is the zero contour.
\label{fig:egs}}
\end{figure}

In~\Fig{fig:egs} (a-c) we present plots of the Wigner function for three different superposition states using~\Eq{highj}.
In comparison to the Wigner function previously defined~\cite{PhysRevA.49.4101,PhysRevA.86.062117} the shape of the functions are quantitatively different; however, these functions do visualize quantum interference in the states in a similar manner.
The advantage of this approach is that the Wigner function can be obtained without a multipole expansion that can be problematic to do for such systems.

While the previous Wigner function is useful for some physical systems, it is inadequate to represent more general spin systems.
To represent the full dynamics of such systems, we need to employ a different symmetry to construct a Wigner function.
One particular general spin system of interest is a multiqubit system, which is a special case of a more general ensemble of qudits~\cite{BillMunro1}.
Although it is possible to imbed the high-$j$ $\SU{2}$ symmetry into the appropriate $\SU{\Dim}$ group representation of the entire Hilbert space of a multiqubit system and generate Wigner functions using~\Eq{highj} (see~\cite{Klimov-1008.2920}), the resulting Wigner function is fully dependent on the labeling of the basis states.
To correct for this, we employ a rotation of the form $\SU{2}\otimes \cdots \otimes \SU{2}$. 
More precisely, for $k$ qubits, we have $n_i=2$ and $d_i=2$ for all $k$, allowing us to define the total rotation operator as: $\hat{\mathbb{U}}=\bigotimes_i^k \hat{U}_2^{[2]}(\theta_i,\phi_i,\Phi_i)_i=\bigotimes_i^k e^{\ui \hat{\sigma}_{z_{i}} \phi_i}e^{\ui \hat{\sigma}_{y_{{i}}} \theta_i}e^{\ui \hat{\sigma}_{z_{i}} \Phi_i}$.
Doing this we obtain
\ba
\label{susu}
\OperBare^{[2^{k}]}(\{\theta_i,\phi_i\}) &=& \frac{1}{2^k}
	\left\{\bigotimes_i^k (\OpU_2^{[2]})_i\right\}
	\hat{\Pi}^{[2^{k}]}
	\left\{\bigotimes_i^k (\OpU_2^{[2]})_i^\dag\right\}, \nonumber \\
\hat{\Pi}^{[2^{k}]} &=& 
	\Id^{[2^{k}]} - 
	\mathcal{N}(2^{k}) \hat{\Lambda}_{2^{2k}-1},
\ea 
where $\mathcal{N}(2^{k})=\sqrt{{(2^k+1)(2^k)(2^k-1)}/{2}}$ (assuming the appropriate Haar measure representation) as well as noting that, once again, the $\Phi_i$'s make no contribution. 

As the number of parameters $\{\theta_i,\phi_i\}$ of~\Eq{susu} scales with the number of qubits/atoms/spins it becomes harder to visualize. 
However, we still can capture the nature of the corresponding state by taking slices of its Wigner function, for instance, by setting $\theta_i=\theta_j$ and $\phi_i=\phi_j$ for all $i,j$.
In~\Fig{fig:egs} (d-f) we show such slicing for all $i,j$ for a selection of states that are usually mapped onto the respective spin states shown in~\Fig{fig:egs} (a-c). 

It is interesting to note that if we write $\OpA = \OpV \OpA_0 \OpV^\dag$ where $\OpV$ is some unitary operator then, in general,
\bel{easy}
\Func{\OpA}=\Trace{ \OpV \OpA_0 \OpV^\dag \, \Oper}=\Trace{  \OpA_0\, \tilde{\Delta}(\Param)},
\ee
where we have a new, rotated kernel $\tilde{\Delta}(\Param)=\OpV^\dag  \Oper \OpV$.
Then, if, for example $\OpA=\DO$ and $\OpV$ is the evolution operator, or a set of quantum gate operations, this expression can lead to an efficient way of computing the Wigner function for a dynamical process or an algorithm as $\OpV^\dag  \Oper \OpV$.
\begin{figure}[!tb]
\includegraphics[keepaspectratio=true,width=0.45\linewidth]{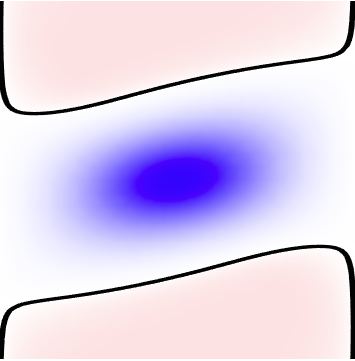}
\hfill
{%
\setlength{\fboxsep}{0pt}%
\setlength{\fboxrule}{1pt}%
\fcolorbox{gray}{white}{\includegraphics[keepaspectratio=true,width=0.45\linewidth]{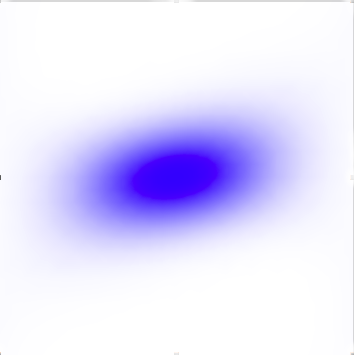}}
}
\caption{Comparison of the Wigner (left) and $Q$ (right) functions for states generated by one axis squeezing of a set of two-level systems. 
The initial state is $\ket{+_6}:=\bigotimes_{i=1}^6 \big(\ket{j=\Half,m=\Half}_i+\ket{j=\Half,m=-\Half}_i \big)$ 
and the Hamiltonian is $\OpH=\left[\bigoplus_i (\hat{\sigma}_z)_i\right]^2$. 
We calculate the Wigner function using~\Eq{easy} using $\OperBare^{[2^{k}]}(\{\theta_i,\phi_i\})$, $\OpA=\OP{+_6}{+_6}$ and $\OpV=\exp(-\ui\OpH t)$ where $t=\pi/125$. 
$Q$ is calculated in the usual way using the natural $\theta$, $\phi$ parametrization of spin coherent states (see~\cite{Nemoto2000} for details). 
As in~\Fig{fig:egs} (d-f), we have taken the slice $\theta_i=\theta$ (as the ordinate) and $\phi_i=\phi$ (as the abscissa).
In both figures we clearly see squeezing, but in the Wigner function we also see negative volume indicating the underlying quantum nature of the states. 
Note, unlike in~\Fig{fig:egs} (d-f), here $\phi$ ranges from $-\pi/2$ to $\pi/2$.
For all plots, blue is positive and red negative and black is the zero contour.
\label{Fig:sq}}
\end{figure} 
An example of the utility of this approach is shown in~\Fig{Fig:sq} where we have applied this method to show squeezing in a set of spins using $\OperBare^{[2^{k}]}(\{\theta_i,\phi_i\})$ for a toy model of one-axis twisting. 

Lastly, we can extend our Wigner function representation to even more spin system symmetries. 
If we set $k = 1$, $n_1 = N$, and $d_1 = D$ we generate the rotational operator $\hat{\mathbb{U}}=\OpU_N^{[D]}$ representing a general $D$-dimensional quantum system or qudit with $\SU{N}$ symmetry (for operator formalism see~\cite{Tilma2}; for coherent state formalism see~\cite{Nemoto2000}). 
The kernel, following our Haar measure requirements, is then 
\ba
\label{TheKT}
\OperBare^{[D]}(\boldsymbol{\theta},\boldsymbol{\phi})&=&\frac{1}{D}
	\OpU_N^{[D]}
	\hat{\Pi}^{[D]}
	(\OpU_N^{[D]})^\dag \nonumber \\
\hat{\Pi}^{[D]}&=&
	\Id^{[D]} -
	\mathcal{N}(D)
	\hat{\Lambda}_{D^{2}-1}
\ea
where $\boldsymbol{\theta}=[\theta_1,\theta_2,\ldots,\theta_{N-1}]$ and $\boldsymbol{\phi}=[\phi_1,\phi_2,\ldots,\phi_{N-1}]$ with $\mathcal{N}(D)=\sqrt{{(D+1)(D)(D-1)}/{2}}$.
Using $ \OpU_N^{[D]}$ with $D = N$ from~\cite{Tilma2}, the above function is then identical to the $\SU{N}$ coherent state-based Wigner function of~\cite{TilmaKae1}.
This allows us to consider the dynamics of a set of $k$ qudits as a mapping onto the dynamics of a coherent state in $\SU{N^k}$, which is a form of holographic principle that reminds us of conformal field theories, by setting the rotation operator to be $\hat{\mathbb{U} }= \bigotimes_i^k (\OpU_N^{[N]})_i$.  
For example, if $N=3$, we generate the kernel for a set of qutrits whose dynamics can be mapped onto that of a coherent state in $\SU{3^k}$. 
Construction of the associated Wigner function proceeds in exactly the same way as before. 
Obviously this can be generalized.
This leads us to propose that operators with other Lie group symmetries, such as SO$(N)$, could be used if we have a reason to believe such symmetries describe the underlying physics of the system.

To conclude, we have shown a general method for constructing Wigner functions using the symmetries contained within the Special Unitary (SU) group.  
This approach allows us to construct and explicitly derive the Wigner functions for arbitrary spin systems. 
Furthermore, as Wigner functions of composite systems can be generated by a kernel that is the tensor product of its components~\cite{Luis-495302} combining existing methods with those presented here provides a mechanism to define Wigner functions for arbitrary quantum systems.
As our ability to quantum coherently control a physical system has been rapidly improving, we can anticipate a large quantum system to be experimentally realized in the relatively near future, and hence we should note that this formalism is numerically, computationally, and experimentally friendly (the Wigner function is the expectation value of a displaced parity operator~\cite{PhysRevA.50.4488} or, equivalently, the expectation value of a parity operator for a state rotated in the opposite direction).  
Lastly, because of the usefulness of the SU group in theoretical physics, this formalism should help generate usable Wigner functions for high-spin $\SU{N}$ systems that are important in theoretical studies of quantum gravity, string theory, and other extensions to quantum mechanics~\cite{1207.1262}. 

We would like to thank Shane Dooley, Emi Yukawa, and Michael Hanks for interesting and informative discussions. 
KN acknowledges support from the MEXT Grant-in-Aid for Scientific Research on Innovative Areas ``Science of hybrid quantum systems'' Grant Number 15H05870 and JSPS KAKENHI Grant Number 25220601. 
Lastly, TT and MJE contributed equally to this work.

%
%
\bibliographystyle{apsrev4-1}
\bibliography{refs}  

\end{document}